\documentclass[aps,prb,twocolumn,superscriptaddress, show pacs]{revtex4-1}
\usepackage{bm, amsmath}
\usepackage{graphicx}
\usepackage{color}
\usepackage{epstopdf}
\usepackage{graphicx}
\usepackage{dcolumn}
\usepackage{bm}
\usepackage{subfigure}

\begin{document}

\title{Toggle-switch-like crossover between two types of isolated skyrmions within the conical phase of cubic helimagnets}


\author{A. O. Leonov}
\thanks{leonov@hiroshima-u.ac.jp}
\affiliation{Chirality Research Center, Hiroshima University, Higashi-Hiroshima, Hiroshima 739-8526, Japan}
\affiliation{Department of Chemistry, Faculty of Science, Hiroshima University Kagamiyama, Higashi Hiroshima, Hiroshima 739-8526, Japan}
\affiliation{IFW Dresden, Postfach 270016, D-01171 Dresden, Germany} 

\author{A. N. Bogdanov}
\affiliation{Chirality Research Center, Hiroshima University, Higashi-Hiroshima, Hiroshima 739-8526, Japan}
\affiliation{Department of Chemistry, Faculty of Science, Hiroshima University Kagamiyama, Higashi Hiroshima, Hiroshima 739-8526, Japan}
\affiliation{IFW Dresden, Postfach 270016, D-01171 Dresden, Germany}

\author{K. Inoue}
\affiliation{Chirality Research Center, Hiroshima University, Higashi-Hiroshima, 
Hiroshima 739-8526, Japan}
\affiliation{Department of Chemistry, Faculty of Science, Hiroshima University Kagamiyama, Higashi Hiroshima, Hiroshima 739-8526, Japan}

\date{\today}

\begin{abstract}
{We investigate the field-induced crossover between two types of isolated skyrmions that exist within the conical phase of cubic helimagnets and orient themselves either along or perpendicular to the field.   
Such a crossover takes place for the same value of the  field, at which the closely packed skyrmion lattice was predicted to stabilize in the A-phase region. 
The clusters and a skyrmion lattice comprised by the skyrmions perpendicular to the field, however, are unfavorable and lose their stability as compared with the skyrmions parallel to the field. 
We also followed transformation of perpendicular skyrmions into pairs of merons that rupture the helical state. 
An attractive interactions between different types of isolated skyrmions make it feasible to construct complex cluster states with the cubic arrangement of skyrmions. 
}
\end{abstract}

         
\maketitle


\textit{1. Introduction.} Magnetic chiral skyrmions are particle-like topological solitons with complex non-coplanar spin structure \cite{JMMM94,Romming13,LeonovNJP16,review} stabilized in noncentrosymmetric magnetic materials by specific Dzyaloshinskii-Moriya interactions (DMI) \cite{Dz64}.
DMI provides a unique stabilization mechanism, protecting  localized states from radial instability \cite{JMMM94,LeonovNJP16} and overcoming the constraints of the Hobart-Derrick theorem \cite{solitons}.
That is why noncentrosymmetric magnets and other chiral condensed matter systems (in particular, chiral liquid crystals \cite{Oswald}) are of special interest in fundamental physics and mathematics as a particular class of materials where skyrmions can exist \cite{Nature06,Melcher14}.

\begin{figure}
\includegraphics[width=0.95\columnwidth]{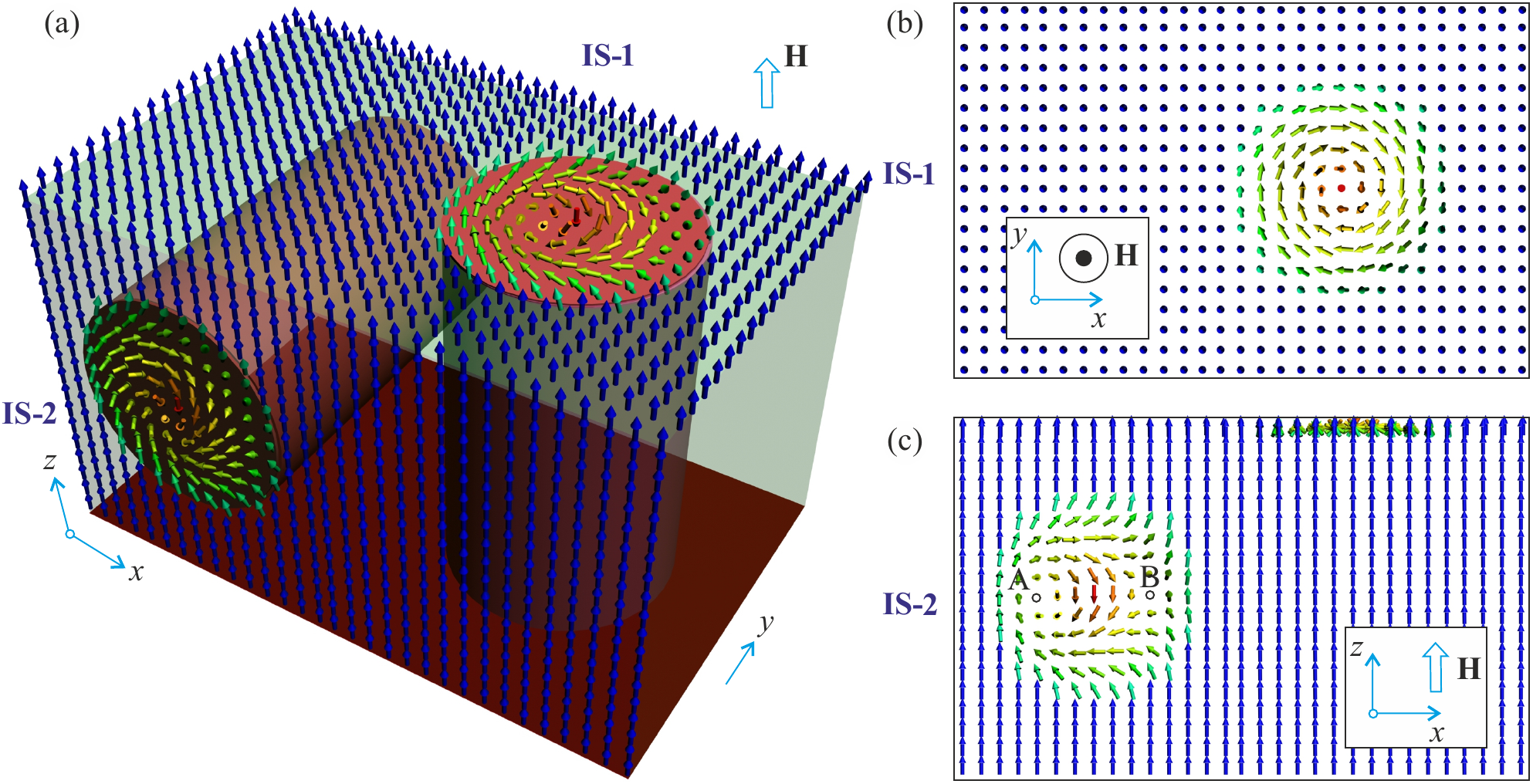}
\caption{ 
(color online) (a) The spin texture of two types of isolated skyrmions oriented either along the field (IS-1) or perpendicular to it (IS-2). 
While IS-1 exhibits an axisymmetric shape (b), IS-2 acquires a non-axisymmetric shape with the circular core section centered around the point A and the anti-skyrmion-like crescent centered around the point B.
\label{art} 
}
\end{figure}

The current interest of skyrmionics is focused on axisymmetric skyrmions within the saturated ferromagnetic states of non-centrosymmetric magnets \cite{LeonovNJP16}. 
Axes of such skyrmions are co-aligned with an applied magnetic field, i.e.  the magnetization in the center is opposite to the field and gradually rotates to the field-saturated state at the outskirt, as visualized in Figs. \ref{art} (a),  (b).
Recently, skyrmion lattice states (SkL-1) and isolated skyrmions (IS-1) of such a type were discovered in bulk crystals of chiral magnets near the magnetic ordering temperatures \cite{Muehlbauer09,Wilhelm11,Kezsmarki15} and in nanostructures with confined geometries over larger temperature regions\cite{Yu10,Yu11,Du15,Liang15}.
The internal structure of such axisymmetric skyrmions, generally characterized by repulsive skyrmion-skyrmion interaction, has been thoroughly investigated theoretically \cite{JMMM94,LeonovNJP16} and experimentally by spin-polarized scanning tunneling microscopy in PdFe bilayers with induced Dzyaloshinskii-Moriya interactions and strong easy-axis anisotropy \cite{Romming13,Romming15}.
%

In Ref. \onlinecite{Leonov17} it was shown, that the saturated state may accommodate another type of isolated skyrmions with their axes perpendicular to field (IS-2).
Such skyrmions are forced to develop  non-axisymmetric shape in order to match the spin pattern of the homogeneous state and to preserve their topological charge $Q = 1$ (Figs. \ref{art} (a), (c)).
Starting from the left side of the depicted skyrmion in the plane $xz$ (Figs. \ref{art} (c)), the magnetization makes the full turn from $0$ to $2 \pi$: it passes through points $A$ and $B$ with the horizontal orientation of the magnetization, $\theta=\pi/2$ and $3\pi/2$, which can be regarded as centers of a circular skyrmion core and an anti-skyrmion crescent, respectively.
Unlike the repulsive axisymmetric skyrmions IS-1, such non-axisymmetric skyrmions IS-2 exhibit anisotropic inter-skyrmion potential \cite{Leonov17}.
Depending on the relative orientation of the two individual skyrmions, this potential can be attractive, leading
to the formation of biskyrmion or multiskyrmion chains, aligned along the axis connecting points $A$ and $B$, as well as repulsive in the perpendicular direction (i.e. along $z$)\cite{Leonov17}.

\begin{figure}
\includegraphics[width=0.98\columnwidth]{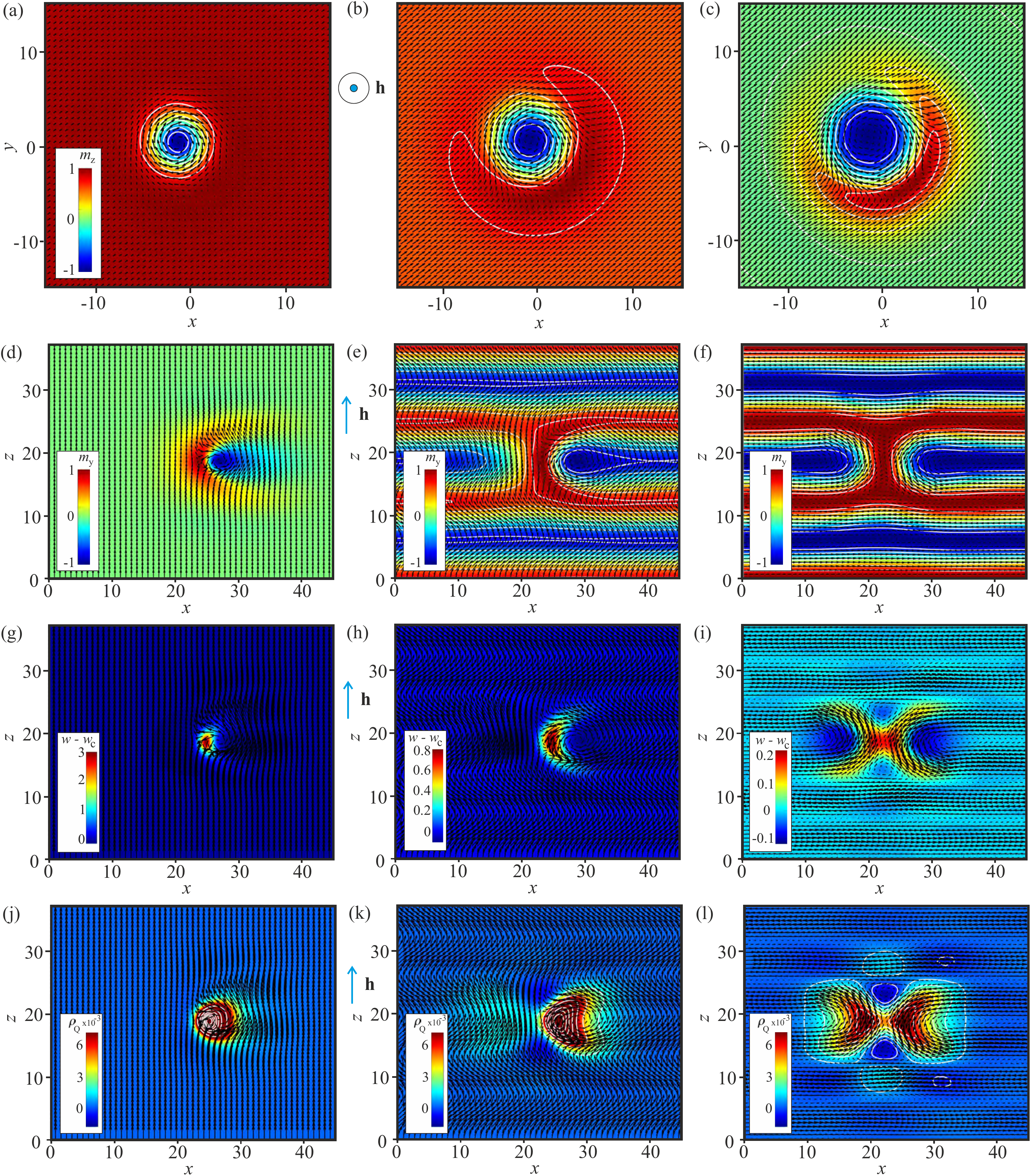}
\caption{
(color online)   
\label{structure} Magnetic structure of IS-1 showing its evolution from the axisymmetric structure for $h=0.6$ (a) to the non-axisymmetric one
for $h=0.3$ (b) and $h=0$ (c). The color plots indicate $z$-component of the magnetization, black arrows are projections of the magnetization
on to the $xy$ plane. The  conical phase has the wave vector co-aligned with the field $h||z$. 
IS-2 for the same orientation of the field 
transforms into a pair of merons that rupture the spiral state. $m_y$-component of the magnetization is shown as color plots in the row from (d) to (f) for different values of the field, while the black arrows indicate projections of the magnetization onto the $xz$ plane. Consequent rows (g), (h), (i)  and (j), (k), (l) exhibit energy density (\ref{density})  counted with respect to the conical phase and the topological charge density, respectively.
}
\end{figure}

The internal spin pattern of isolated skyrmions IS-1 
can also violate the rotational symmetry once placed into the conical phase of bulk cubic helimagnets \cite{LeonovJPCM16}.  
Then, the magnetization distribution in a cross-section $xy$ of IS-1 resembles the structure of skyrmions IS-2. 
The central core region nearly preserves the axial symmetry (Fig. \ref{structure} (a) - (c)), whereas the domain-wall region, connecting the core with the embedding conical state is \textit{asymmetric} \cite{LeonovJPCM16} and acquires a crescent-like shape. This asymmetric profile of the cross-section forms a screw-like modulation along $z$ axis, matching the rotating magnetization of the conical phase. 
These asymmetric skyrmions were shown to exhibit an isotropic \textit{attractive} skyrmion-skyrmion interaction \cite{LeonovJPCM16}.
They are also believed to underlie precursor phenomena (e.g., the A-phase) near the ordering temperatures in
chiral B20 magnets (MnSi \cite{Muehlbauer09}, FeGe \cite{Wilhelm11}) \cite{LeonovJPCM16} and to have prospects in spintronics as an alternative to the common axisymmetric skyrmions \cite{LeonovAPL16}.

In the present paper, we consider a field-driven transformation of isolated skyrmions IS-2 
when the surrounding saturated state turns into the conical state with the lowering magnetic field.
We show that an isolated skyrmion IS-2 transforms into a pair of merons  with equally distributed topological charge $Q=1/2$, whereas a couple of attracting skyrmions IS-2 -- into one axisymmetric skyrmion and a pair of merons, since
such a configuration is energetically more favorable than two separated pairs of merons and stems from an attractive inter-skyrmion interaction as was pointed out in Refs. \onlinecite{Muller,Leonov17}.
We derive regular solutions for IS-2 and prove that for some threshold-field they become a lower metastable solution as compared with IS-1.
%
%
At the same time, the energy gain by clustering for IS-2 is lower than the gain by the same clustering of IS-1.
As a result, a skyrmion lattice SkL-2 is highly metastable and loses its stability in an applied magnetic field.


\textbf{2.} \textit{Model.} The standard model for magnetic states in bulk cubic non-centrosymmetric ferromagnets is based on the energy density functional \cite{Dz64,Bak80}
\begin{equation}
w =A\,(\mathbf{grad}\,\mathbf{m})^2 + D\,\mathbf{m}\cdot \mathrm{rot}\,\mathbf{m} -\mu_0 \,M  \mathbf{m} \cdot \mathbf{H},
\label{density}
\end{equation}
including the principal interactions essential to stabilize modulated states:  the exchange stiffness with constant $A$,  Dzyaloshinskii-Moriya  coupling energy with constant $D$, and the Zeeman energy; $\mathbf{m}= (\sin\theta\cos\psi;\sin\theta\sin\psi;\cos\theta)$  is the unity vector along the magnetization vector  $\mathbf{M} = \mathbf{m} M$, and $\mathbf{H}$ is the magnetic field applied exclusively along $z-$ axis.
%

The solutions for particle-like skyrmions of both introduced types IS-1 and IS-2 (Fig. \ref{art}) are derived by the Euler equations for energy functional (\ref{density}) together with the Maxwell equations. 
The skyrmion solutions depend only on the value of the applied magnetic field, $h = H/H_D$, where $\mu_0 H_D = D^2/(A M)$ is the \textit{saturation field} \cite{Bak80,JMMM94}.

The twisting magnetization $\mathbf{m}$ in the skyrmions also matches boundary conditions imposed by the surrounding conical phase. 
The equilibrium parameters for this cone phase  are expressed in the analytical form \cite{Bak80}  as:
\begin{eqnarray}
\theta_c = \arccos \left(2H/H_D \right), \, 
 \psi_c =   z/2L_D, \, \, 
\label{cone}
\end{eqnarray}
Above the critical field $H=0.5H_D$,  the cone phase transforms into the saturated state with $\theta = 0$. 
$L_D = A/|D|$ is the characteristic length unit of the modulated states.


\textit{3. The internal structure of IS-1 and IS-2.}
%
Evolution of two types of ISs with the decreasing magnetic field $H<H_D$ is shown in Fig. \ref{structure}:

(a) IS-1,  perfectly fitting into the magnetically saturated background for higher magnetic fields (Fig. \ref{structure} (a), $h=0.6$), gradually develops a non-axisymmetric shape with the onset of the conical phase (Fig. \ref{structure} (b), (c), $h=0.3,\,0$). 
Unlike the axisymmetric skyrmions with the essentially two-dimensional structure and their axes perfectly aligned along the field (Fig. \ref{structure} (a)), the non-axisymmetric skyrmions (Fig. \ref{structure} (b), (c)) become truly three-dimensional and are inhomogeneous along $z-$axis in a sense that the skyrmion axes now circumscribe a screw-like trajectory and the magnetization pattern in the basal $xy$ plane follows the magnetization in the conical phase at each cross-section (see Supplementary Material in Ref. \onlinecite{Loudon18}).
Since these non-axisymmetric skyrmions are incompatible with the conical phase, they are surrounded by a circle-like "shell" -- a transitional region with the positive energy that underlies an attractive inter-skyrmion potential and leads to the skyrmion cluster formation (see Refs. \onlinecite{LeonovJPCM16,LeonovAPL16,Loudon18} for extensive details on the structure of these IS-1 and their experimental observation in thin layers of a cubic helimagnet Cu$_2$OSeO$_3$).
The properties of repulsive axisymmetric skyrmions within the saturated state (Fig. \ref{structure} (a))  were theoretically investigated in Refs. \onlinecite{JMMM94,LeonovNJP16} and experimentally examined in nanolayers of chiral ferromagnets \cite{LeonovNJP16,Romming15}.
%

(b) At higher magnetic fields (Fig. \ref{structure} (d), $h=0.6$), IS-2 has a non-axisymmetric structure in the plane $xz$ that is homogeneously replicated along $y$-direction. 
The asymmetric magnetic structure of these skyrmions is associated with an intricate pattern of internal characteristics like the energy density (Fig. \ref{structure} (g))  or the topological charge distribution (Fig. \ref{structure} (j)) \cite{Leonov17,Lin2015}. 
Consequently due to the positive and negative asymptotics of the magnetization  with respect to the surrounding state, such skyrmions develop an attractive interaction in the basal $xy$ plane and a repulsive interaction along $z$-axis \cite{Leonov17}.
The stability of these asymmetric skyrmions was theoretically investigated in Ref. \onlinecite{Leonov17} with the clear directions for their experimental studies in a polar easy-plane magnet GaV$_4$Se$_8$ \cite{Ruff15}. 
Note however, the structure of skyrmions stabilized by the easy-plane anisotropy (even for zero magnetic field) \cite{Leonov17} exhibits smooth rotation of the magnetization as compared with  IS-2 that are strongly localized by the unidirectional magnetic field (with zero anisotropy): in spite of the seeming similarity between the magnetization patterns, these skyrmions   bear different patterns of the topological charge or energy density distribution (see Ref. \onlinecite{Leonov17} for details).

With the onset of the conical phase, IS-2 undergoes an elliptical instability (Fig. \ref{structure} (e), (h), (k), $h=0.3$). 
Still, the characteristic distributions of the topological charge and energy densities remain localized what makes IS-2 countable particles (Fig. \ref{structure} (h), (k)).
For $h=0$, the transformation of an IS-2 into a pair of merons (representing a rupture of a spiral state) with the topological charge $Q=1/2$ each is complete (Fig. \ref{structure} (f), (i), (l)). 
The properties of such skyrmions and merons within the helical background were recently investigated in Refs. \onlinecite{Muller,Ezawa}. 
It was argued that they behave as if they were free particles, and the helical background provides natural one-dimensional channels along which skyrmions and merons can move rapidly \cite{Muller}.


\begin{figure}
\includegraphics[width=0.98\columnwidth]{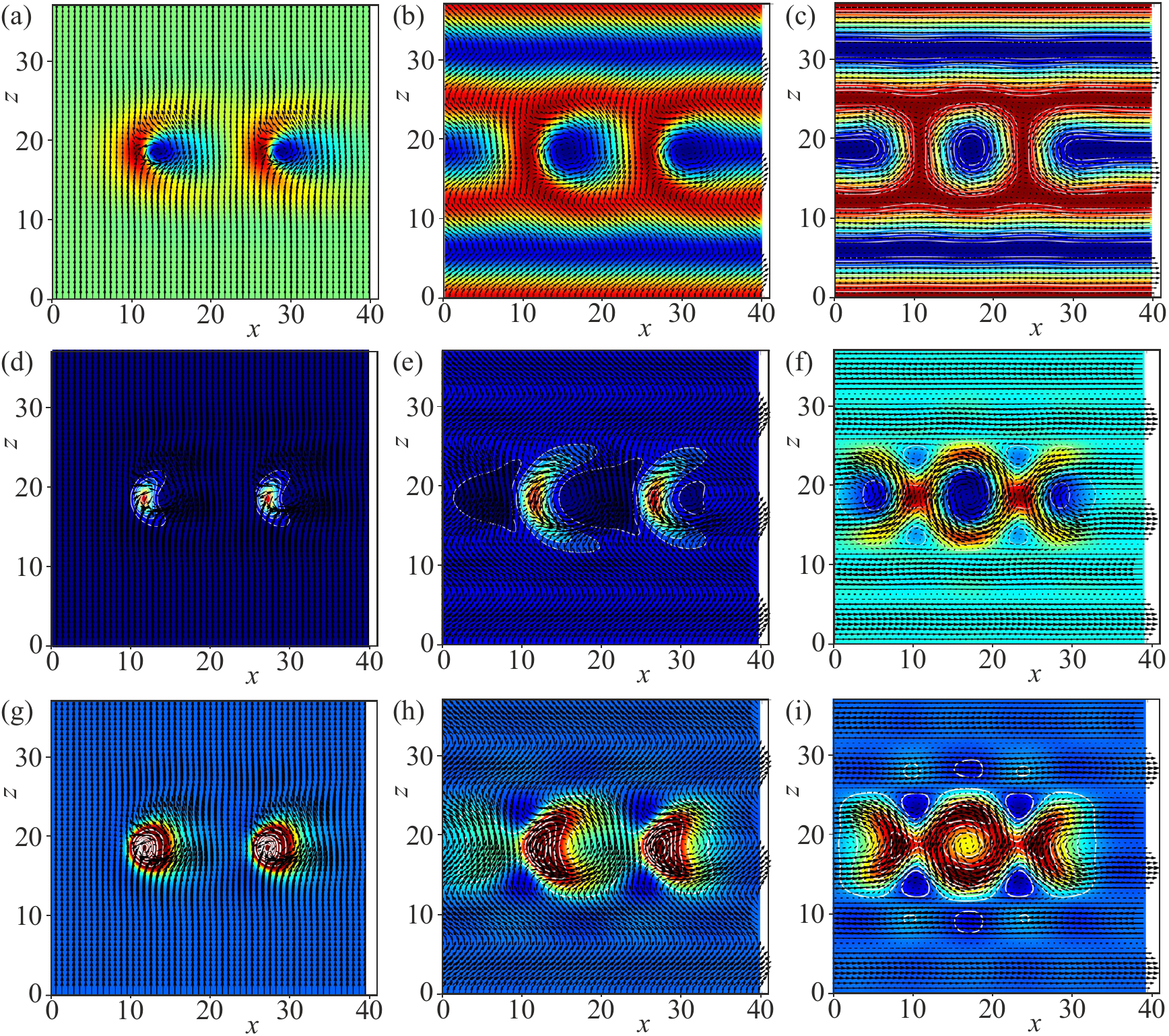}
\caption{
(color online) The same characteristics as plotted for IS-2 with $Q=1$, but for a coupled pair of IS-2 with $Q=2$: the first row exhibits 
 $m_y$-component of the magnetization, the second row - energy density, and the third row - the topological charge density for different values of the applied field.  
\label{pair}
}
\end{figure}

\begin{figure}
\includegraphics[width=0.98\columnwidth]{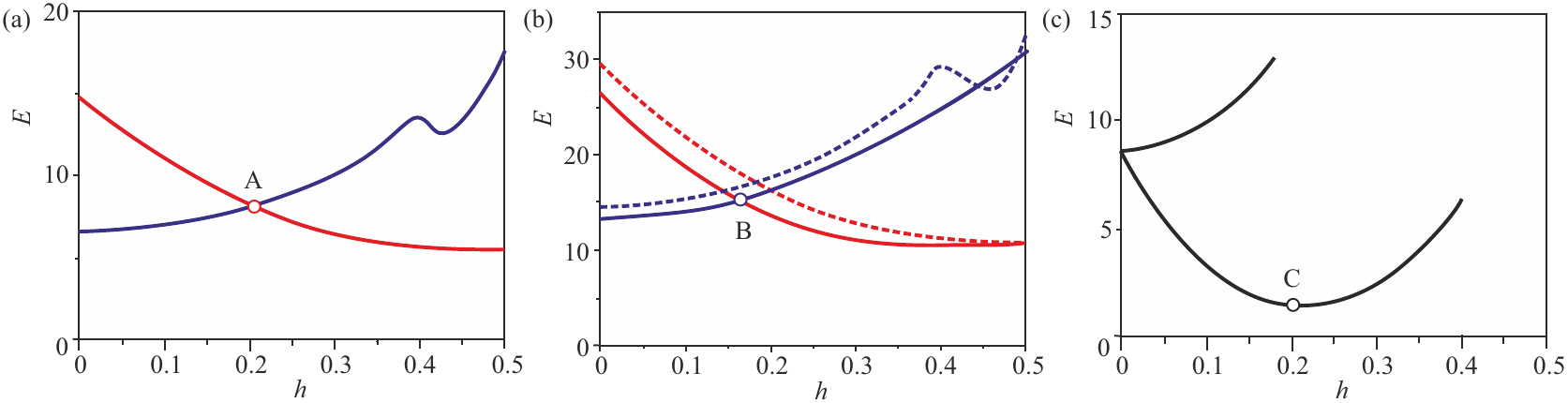}
\caption{
(color online) Energy of IS-1 (red curves) and IS-2 (blue curves) plotted in dependence on the applied magnetic field for two configurations, with $Q=1$ (a) (as depicted in Fig. \ref{structure}) and $Q=2$ (b) (as depicted in Fig. \ref{pair}). The crossover between two types of isolated skyrmions (point $A$ in (a)) is shifted towards lower fields with adding more skyrmions into skyrmion clusters (point $B$ in (b) for two skyrmions in a cluster).  Dashed and solid lines show the energies of remote skyrmions and coupled skyrmion pairs, correspondingly. (c) Energy densities of SkL-1 and SkL-2. Only SkL-1 can be stabilized by additional anisotropic interactions around the critical field, $h=0.2$ (point $C$).
\label{energy}
}
\end{figure}

\textit{4. Field-driven crossover between IS-1 and IS-2.}
%
Since IS-1 perfectly blends into the  homogeneous state saturated along the field (Fig. \ref{structure} (a), $h=0.6$) whereas IS-2 - into the helicoid (Fig. \ref{structure} (f), $h=0$),  a crossover between two states takes place for an intermediate value of the field, $h\approx 0.2$ (red and blue solid lines with an intersection point $A$ in Fig. \ref{energy} (a)).
We dubbed this crossover a toggle-switch-like transition since it is associated with the flip of a skyrmion axis by $90^\circ$   
%
- from the orientation along $z-$axis to the in-plane $xy$ orientation.

Due to the transient region (shell) formed by each type of a skyrmion with respect to the conical phase, one should expect an attractive interaction also between IS-1 and IS-2 (a skyrmion-skyrmion potential will be computed elsewhere). The clusters consisting from the mutually perpendicular skyrmions will then exhibit an entangled pattern since IS-2 do not have any preferable in-plane orientation within an isotropic model (\ref{density}). 
On the contrary, an additional cubic anisotropy may align IS-2 along easy  directions, e.g., $<100>$, thus a phase reminiscent a 
uniaxial model for the Blue Phase II in chiral liquid crystals with symmetry $O_2$ may occur (see Fig. B.VIII.12 in Ref. \onlinecite{Oswald}).
%

The IS-1--IS-2 crossover field is shifted towards the lower fields  when pairs of corresponding skyrmions are considered (red and blue solid lines with an intersection point $B$ in Fig. \ref{energy} (b)). 
The reason lies in the following: the energy profit due to the clustering is much higher for IS-1. 
Dashed red and blue lines in Fig. \ref{energy} (b) show energies of skyrmion pairs at an infinite distance from each other; by the gap with the solid lines of the corresponding color one could estimate the energy gain.

The SkL-2 as an ultimate outcome of clustering has higher energy as compared with SkL-1 and easily loses its stability in an applied magnetic field (black lines in Fig. \ref{energy} (c)).

The IS-1--IS-2 crossover field (point $A$ in Fig. \ref{energy} (a)) within a numerical error equals the field at which the difference between the energy densities of the hexagonal SkL-1 and the cone phase reaches its minimum (point $C$ in Fig. \ref{energy} (c)). 
Thus in Ref. \onlinecite{Wilson14}, it was suggested that the SkL-1 could be stabilized with respect to the cones by additional anisotropic energy contributions (e.g., by the uniaxial anisotropy of the easy-axis type \cite{Butenko10})  exactly around this field value.
With our new insight, we may add that although the hexagonal SkL-2 has much higher energy as compared with SkL-1, separate IS-1--IS-1, IS-2--IS-2, and IS-1--IS-2 clusters 
glued together owing to the mutual attracting interaction may create a frozen metastable state which may become paramount within A-phases of bulk cubic helimagnets near the ordering temperature (e.g., in B20 magnets MnSi \cite{Muehlbauer09} and FeGe \cite{Wilhelm11}).


\textit{5. IS-2--IS-2 interaction.}
According to Fig. \ref{energy} (b) for $H/H_D=0.5$ (blue solid and dotted lines), the energy of two remote skyrmions (Fig. \ref{structure} (e)) is higher than the energy of a skyrmion cluster with two  skyrmions in the plane $xy$ (Fig. \ref{pair} (b)).  Hence, we talk about an attractive inter-skyrmion potential and formed skyrmion chains. Such a result complies with Ref. \onlinecite{Leonov17}. 
With the decreasing magnetic field, $H<0.5H_D$, however, we do not talk about skyrmion-skyrmion interaction, since the skyrmions and merons become positionally confined within one spiral period and one considers different configurational arrangements  of skyrmions and merons (Fig. \ref{structure} or Fig. \ref{pair}).

According to the results of Ref. \onlinecite{Muller}, the configuration with two merons nearing a skyrmion (Fig. \ref{pair} (c)) is energetically more favourable than two distant merons (Fig. \ref{structure} (f)).
In that geometry \cite{Muller} the field is applied along $y$-direction, and the energy of a meron pair becomes negative with respect to the helical background above the critical field $h_y\approx 0.16$: it gives start to the helicoid-SkL first-order phase transition.
In the present geometry with $h||z$, we find additional intersection points between blue dotted and solid lines that occur at $H<H_D$: they indicate the crossover between these two configurational states. 
Moreover, IS-2 bears a positive energy density (Fig. \ref{energy} (a)) in the whole field range and thus represents a metastable state within the conical background. 
%

%

Such a study, however, requires an extensive investigation (will be done elsewhere) since contradicting results are reported so far in the literature.
In particular in Ref. \onlinecite{Lin2015}, it was shown that the skewed skyrmions IS-2 have anisotropic inter-skyrmion interaction, i.e. the interaction energy between two skyrmions depends on their relative orientation.
Nevertheless, the skyrmion-skyrmion interaction remains repulsive with the strongest repulsion between skyrmions in the direction where they are elongated \cite{Lin2015}.
In Ref. \onlinecite{MullerThesis} however, the IS-2--IS-2 interaction was reported to be oscillating along the perpendicular $z$-direction with alternating minima and maxima, whereas it is strongly repulsive in the $xy$ plane. 
In Ref. \onlinecite{Leonov17}, the interaction was claimed to be attracting in the plane $xy$ and repulsive along the $z$-direction.

\textit{6. Conclusions. }
In conclusion, we found a type of isolated skyrmions that emerge in cubic helimagnets and orient their axes perpendicular to the field -- IS-2. 
These skyrmionic states are characterized by an asymmetric shape that also stipulates their attracting interaction within the field-saturated state. 
With the onset of the conical phase with its $q$-vector along the field, IS-2 gradually transform into pairs of merons.
The close packing of merons is energetically preferable almost in the whole field interval. 
We also discovered a crossover between IS-2 and IS-1 that would occur just in the middle of the A-phase region.
In order to fully explore the characteristics and functionalities of IS-2, their internal structure should be studied experimentally, as was done for the axisymmetric individual skyrmions within polarized FM states \cite{LeonovNJP16}: in the same geometry as in Ref. \onlinecite{LeonovNJP16}, one should apply the field in the film plane.

\section{Acknowledgements}

The authors are grateful to Ivan Smalyukh, Jun-ichiro Ohe and Istvan Kezsmarki for useful discussions. This work was funded by JSPS Core-to-Core Program, Advanced Research Networks (Japan) and JSPS Grant-in-Aid for Research Activity Start-up 17H06889. AOL thanks Ulrike Nitzsche for
technical assistance.


\begin{thebibliography} {99}

\bibitem{JMMM94} A. N. Bogdanov and A. Hubert, J. Magn. Magn. Mater. \textbf{138}, 255 (1994); \textbf{195}, 182 (1999).


\bibitem{LeonovNJP16} A. O. Leonov, T. L. Monchesky, N. Romming, A. Kubetzka, A. N. Bogdanov, and R. Wiesendanger,
New J. of Phys. \textbf{18}, 065003 (2016).

\bibitem{review} U. K. R\"o\ss ler \textit{et al.}, 
J. Phys. Conf. Ser. \textbf{303}, 012105 (2011).

\bibitem{Romming13} N. Romming \textit{et al.}, 
 Science \textbf{341}, 636 (2013).

\bibitem{Dz64}  I. E.  Dzyaloshinskii, Sov. Phys. JETP {\textbf{19}}, 960 (1964).

\bibitem{solitons} R. Rajaraman, \textit{Solitons and Instantons: An Introduction to Solitons and Instantons in Quantum Field Theory} (North-
Holland, Amsterdam, 1982).

\bibitem{Oswald} P. Oswald and P. Pieranski, \textit{Nematic and Cholesteric Liquid Crystals: Concepts and Physical Properties Illustrated by Experiments} (CRC Press, 2005).

\bibitem{Nature06} U. R. R\"o\ss ler \textit{et al.}, 
Nature \textbf{442}, 797 (2006).

\bibitem{Melcher14} C. Melcher,  Proc. R. Soc. A \textbf{470}, 20140394 (2014).

\bibitem{Muehlbauer09} S. M\"uhlbauer  \textit{et al.}, 
Science \textbf{323}, 915–919 (2009).

\bibitem{Wilhelm11} H. Wilhelm \textit{et al.}, 
Phys. Rev. Lett. \textbf{107}, 127203 (2011).

\bibitem{Kezsmarki15} I. Kezsmarki \textit{et al.}, Nat. Mater. \textbf{14}, 1116–1122 (2015).


\bibitem{Yu10} X. Z. Yu \textit{et al.}, 
Nature (London) \textbf{465}, 901 (2010).

\bibitem{Yu11} X. Z. Yu  \textit{et al.}, 
Nat. Mater. \textbf{10}, 106–109 (2011).

\bibitem{Du15} H. Du \textit{et al.}, 
Nat. Commun. \textbf{6}, 7637 (2015).

\bibitem{Liang15} D. Liang \textit{et al.}, 
Nat. Commun. \textbf{6}, 8217 (2015).

\bibitem{Romming15} N. Romming \textit{et al.}, 
Phys. Rev. Lett. 114, 177203 (2015).

\bibitem{Leonov17} A. O. Leonov and I. Kezsmarki, Phys. Rev. B \textbf{96}, 014423 (2017).

\bibitem{LeonovJPCM16} A. O. Leonov, T. L. Monchesky, J. C. Loudon, and A. N. Bogdanov,  J. Phys.: Condens. Matter. \textbf{28}, 35LT01 (2016).

\bibitem{LeonovAPL16} A. O. Leonov, J. C. Loudon, A. N. Bogdanov,  Appl. Phys. Lett. \textbf{109}, 172404 (2016).

\bibitem{Muller} J. M\"uller \textit{et al.}, 
 Phys. Rev. Lett. \textbf{119} 137201 (2017).

\bibitem{Bak80} P. Bak  and M. H. Jensen,  J. Phys.C: Solid State Phys. \textbf{13}, L881 (1980).

\bibitem{Loudon18} J. C. Loudon \textit{et al.}, 
Phys. Rev. B \textbf{97}, 134403 (2018). 

\bibitem{Lin2015} S.-Z. Lin \textit{et al.}, 
Phys. Rev. B \textbf{91}, 224407 (2015).

\bibitem{Ruff15} S. Bordacs \textit{et al.}, 
Sci. Rep. 7: 7584 (2017).  

\bibitem{Ezawa} M. Ezawa, Phys. Rev. B \textbf{83}, 100408 (2011).

\bibitem{Wilson14} M. N. Wilson \textit{et al.}, 
Phys. Rev. B \textbf{89}, 094411 (2014).

\bibitem{Butenko10} A. B. Butenko \textit{et al.}, 
Phys. Rev. B 82, 052403 (2010).

\bibitem{MullerThesis} J. M\"uller, PhD Thesis, University of Cologne (2018).

\end{thebibliography}
\end{document}